\documentclass{an}
\usepackage{graphicx}
\usepackage{times}
\usepackage{fancyhdr}
\newcommand{\HI}{H{\,\small I}}

\begin{document}

\title{Merger-origin of radio galaxies investigated with H I observations}

\author{B.H.C. Emonts\inst{1}, R. Morganti\inst{1,2}, T.A. Oosterloo\inst{1,2}, J.M. van der Hulst\inst{1}, C.N. Tadhunter\inst{3}, G. van Moorsel\inst{4}, J. Holt\inst{3}}
\institute{Kapteyn Astronomical Institute, University of Groningen, PO Box 800, 9700 AV Groningen, The Netherlands
\and
Netherlands Foundation for Research in Astronomy, Postbus 2, 7990 AA Dwingeloo, The Netherlands
\and
Department of Physics and Astronomy, University of Sheffield, Sheffield S3 7RH, UK
\and
National Radio Astronomy Observatory, Socorro, NM 87801, USA}

\date{Received; accepted; published online}

\abstract{We present results of an \HI\ study of a complete sample of nearby radio galaxies. Our goal is to investigate whether merger or interaction events could be at the origin of the radio-AGN activity. Around five of our radio galaxies, hosted mainly by early-type galaxies, we detect extended \HI\ in emission. In most cases this \HI\ is distributed in large (up to 190 kpc) and massive (up to M$_{\rm \HI} \sim 10^{10}$ M$_{\odot}$) disk- or ring-like structures, that show fairly regular rotation around the host galaxy. This suggests that in these systems a major merger did occur, but at least several Gyr ago. For the \HI-rich radio galaxy B2 0648+27 we confirm such a merger origin through the detection of a post-starburst stellar population that dominates the visible light throughout this system. The timescale of the current episode of radio-AGN activity in our \HI-rich radio galaxies is many orders of magnitude smaller than the merger timescales. Therefore the radio-AGN activity either started late in the lifetime of the merger event, or is not directly related to the merger event at all. Another intriguing result is that the \HI-rich ($> 10^{9}$ M$_{\odot}$) radio galaxies in our sample all have compact radio sources, while none of the extended radio sources contain these amounts of extended \HI. This strongly suggests that there is a relation between the size of the radio jet and the presence of large amounts of neutral gas associated with the host galaxy.
}

\correspondence{emonts@astro.rug.nl}

\maketitle

\section{Introduction}
\label{sec:inrtoduction}
 
Merger and interaction events can supply the central regions of galaxies with gas. They even have often been invoked to remove enough angular momentum from this gas so that it can be accreted onto a super-massive central black hole and in that way serve as fuel to drive an Active Galactic Nucleus (AGN). While there are studies that indeed find a correlation between merger/interaction events and AGN activity (e.g. \cite{hec86,wu98,can01}), other studies find no such correlation  (for example \cite{lut98,dun03,gro05}). In this paper we present results of a project aimed at looking for merger or interaction events in a complete sample of {\sl nearby radio galaxies} in order to investigate whether such a correlation exists for this type of AGN.

As a tracer for the merger/interaction events we use \HI\ observations and additional optical long-slit spectra. In a merger involving gas-rich systems, part of the gas serves as fuel to ignite bursts of star formation (e.g. \cite{bar96,mih96,kap05,spr05}), while another part is expelled to form faint but extended structures (tidal-tails, bridges, shells, disks, etc.) that can survive for a long time, even when in the optical kinematic signatures of the merger have already faded away (e.g. \cite{bar02}, \cite{hib96}). From the analysis of the neutral gas we derive the properties of a possible merger event as well as the timescales involved. These are constrained even further by determining the ages of the stellar populations in these radio galaxies from our optical spectra. We compare this with the properties of the radio sources, in order to establish whether or not there is a relation between merger events and AGN activity in these nearby radio galaxies.

\section{Sample selection}
\label{sec:sample}

Our sample consist of 19 northern radio galaxies from the B2-catalog ($F_{\rm 408MHz} \ga 0.2$ Jy) up to a redshift of {\sl z} = 0.04, observed with the Westerbork Synthesis Radio Telescope (WSRT) and the Very Large Array (VLA) in C-configuration. The sample is a {\sl complete sample}, with the restriction that we left out {\sl (1)} sources in dense cluster environments (since here merger features are likely wiped out on relatively short timescales); {\sl (2)} BL-Lac objects (since we do not know their radio properties). The sample consists of 6 compact ($< 15$ kpc) and 13 extended FR-I ($> 15$ kpc) radio sources, which mainly have {\sl early-type host-galaxies}. We do not have any powerful FR-II sources in our sample, since they are found at higher {\sl z}, where it is difficult to observe \HI\ in emission. In addition we observed 2 more compact radio sources that were not in the original sample, although we stress that these two sources are left out of the statistical analysis of the sample. In a future paper we will also present the results of an additional small sample of bright southern radio sources.

\section{Results}
\label{sec:results}

\subsection{Neutral hydrogen gas}
\label{sec:hi} 

\begin{figure*}
\includegraphics[width=17.5cm]{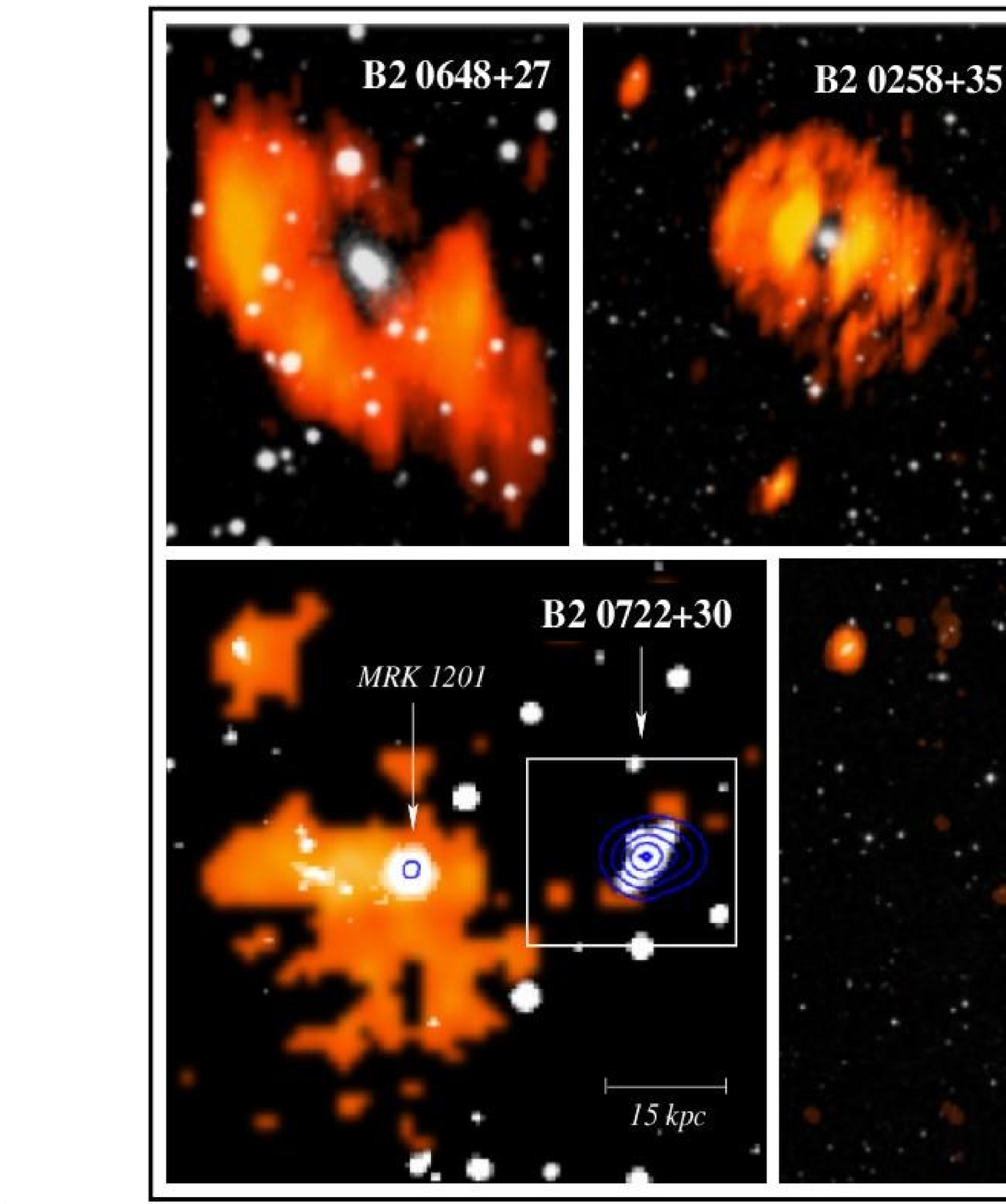}
\caption{0th-moment total intensity maps of the \HI\ emission (red) around five nearby radio galaxies from our sample. Radio continuum of our sample sources is only shown in the cases where the radio source is resolved (blue contours).}
\label{fig:hi}
\end{figure*}

\begin{table}
\caption{{\sl \HI\ around radio galaxies.} Given is the name and - in case the data is from the literature - the reference for the \HI\ mass (+ other related references), the NGC number, the total \HI\ mass detected in emission (sources from the literature have masses corrected for our assumed distance to the galaxy), the size of the \HI\ structure (or distance to the host galaxy for B2 1322+36), and the morphology of the \HI\ structure (D = disk, R = ring, S = shell, B = ''blob'').}
\label{tab:hiradiogalaxies}
\begin{tabular}{lllllll}
$\#$ & Name & NGC & M$_{\rm \HI}$ & D$_{\rm \HI}$ & Morph. \\
 &   &  & (M$_{\odot}$) & (kpc) & \HI \\
\hline
\hline
1 & B2 0258+35         & 1167  & 1.8$\times$10$^{10}$ & 160 & D \\
2 & B2 0648+27         & -  & 7.4$\times$10$^{9}$  & 190 & R \\
3 & B2 0722+30         & -  & 1.5$\times$10$^{8}$  & 15  & D (+B) \\
4 & B2 1217+29$^{\ a\ (b,c)}$ & 4278  & 7.0$\times$10$^{8}$  & 16  & D \\
5 & B2 1322+36         & 5141  & 6.9$\times$10$^{7}$  & 20  & B \\
6 & -                  & 3894   & 2.2$\times$10$^{9}$  & 105 & R \\
\hline
7 & PKS1718-649$^{\ d}$& 6328  & 1.5$\times$10$^{10}$ & 127 & D \\
8 & Cen A$^{\ e\ (f)}$      & 5128  & 6.0$\times$10$^{8}$  & 30  & D+S \\       
\hline
\hline
\end{tabular}

\vspace{1mm} 
References: (a) \cite{lee94} (with D = 14.9 Mpc), (b) \cite{rai81}, (c) Morganti et al. (these proceedings), (d) \cite{ver95}, (e) \cite{sch94} (with D = 3.5 Mpc), (f) \cite{gor90}.
\end{table}

Of our complete sample of 19 galaxies, we detect \HI\ in emission associated with 5 (or $26 \%$) of them, as well as around NGC 3894, which was observed as one of the two additional compact radio sources (Section \ref{sec:sample}). The total intensity images of these \HI\ structures are shown in Figure \ref{fig:hi} (one of our sample sources, B2 1217+29, was already observed by \cite{rai81,lee94}, Morganti et al. - these proceedings). Table \ref{tab:hiradiogalaxies} lists the properties of the \HI\ structures.\footnote{We assume H$_{0}$ = 71 km s$^{-1}$ Mpc$^{-1}$ and use the redshift (unless otherwise indicated) to calculate the \HI\ properties.} For comparison, we also include in this table two other nearby radio galaxies from the literature that are observed in \HI\ emission: Cen A (\cite{sch94}) and PKS 1718-649 (\cite{ver95}). Most of the \HI-rich radio galaxies contain a compact radio source (we will get back to this in Section \ref{sec:radio}).

It is striking that a number of nearby radio galaxies, which are mostly hosted by early-type galaxies, contain vast amounts of \HI, in some cases with masses more than 10$^{10}$ M$_{\odot}$ and spread across structures over 100 kpc in size. We argue that these large \HI\ structures are likely relics of a major merger event. Although there is still a varying degree of asymmetry visible in some of the \HI\ structures, the \HI\ is in most cases located in regular rotating disk- or ring-like structures. This morphology of the \HI\ structures suggests that the merger must have happened at least several Gyr ago, after which the gas had a few galactic orbits time to fall back onto the host galaxy (e.g. \cite{bar02}). B2 1322+36 and B2 0722+30 show only modest amounts of \HI\ ($\sim$ 10$^{8}$ M$_{\odot}$) associated with the radio galaxy. Although for B2 0722+30 most of this \HI\ is again located in a disk, in both these galaxies we detect \HI\ ``blobs'' in the direction of a nearby companion galaxy, possibly tracing a bridge-like structure. From Figure \ref{fig:hi} it is also clear that in particular B2 0722+30 and NGC 3894 are located in an \HI-rich environment of nearby interacting galaxies. 

\HI\ absorption is detected in at least seven of our sample sources. The only source for which the \HI\ is slightly extended in absorption against the radio lobes is B2 1322+36, in the other cases the absorption is unresolved against the central component. This includes the sources found in \HI\ emission (except B2 1217+29, which has no apparent absorption; \cite{lee94}) as well as a few extended sources for which no \HI\ emission was detected. A more detailed study of the \HI\ absorption is necessary to trace the gas in the central region of the radio galaxies. 

In a future paper (Emonts et al. in prep.) we will present a detailed analysis of the \HI\ emission features, investigate the role of the \HI\ environment on the AGN activity, and study the \HI\ absorption properties of our sample sources. 

\subsection{Starbursts}
\label{sec:starburst} 

\begin{figure}
\includegraphics[width=8.4cm]{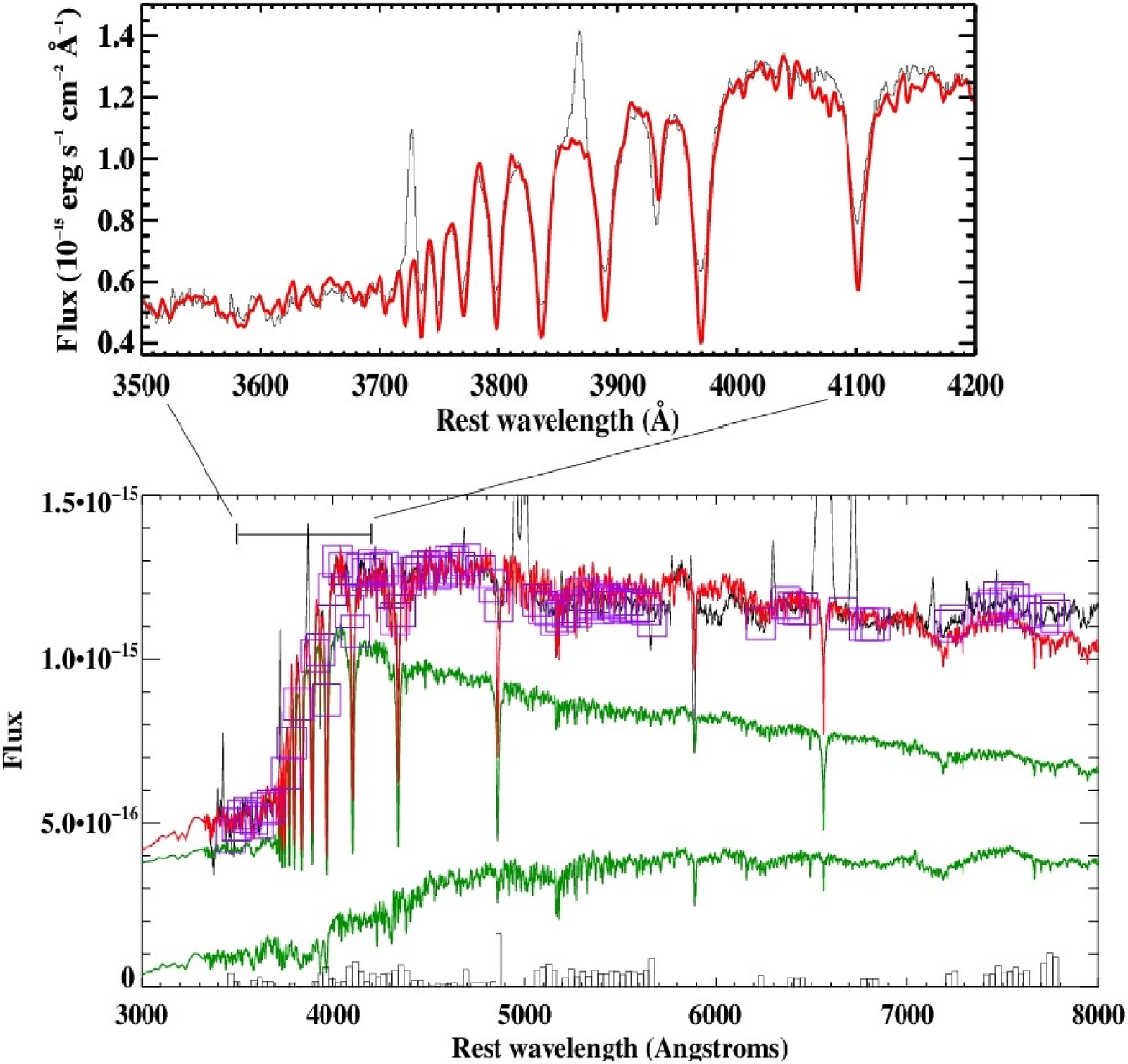}
\caption{Optical spectrum of B2 0648+27 (nucleus). In red is shown the best fit to the spectrum, consisting of both a 12.5 Gyr old stellar population and a 0.2 Gyr young stellar population model with reddening E(B-V) = 0.4 (green). Fitting was done using models of \cite{bru03} across a range of ages for the stellar populations (assumed Salpeter IMF, solar-metallicity instantaneous starburst), and with various reddening. Including a weak power-law component gives a result similar to the best fit shown above without a power-law component. The zoom-in shows the region of Ca{\,\small II} K and H{\small $\delta$}. Also in the off-nuclear regions the light is dominated by a post-starburst stellar population of similar age.}
\label{fig:optical}
\end{figure}
Optical long-slit spectra of our \HI-rich radio galaxies that we took with the William Herschel Telescope (WHT) can be used to determine the stellar populations in these systems by modeling the Spectral Energy Distribution (SED) (see \cite{tad02}, 2005, \cite{wil02}, 2004). In this way we can trace and date possible starbursts that are related to the merger event. The first radio galaxy in our sample for which we did such an analysis is B2 0648+27, as shown in Fig. 2 (Emonts et al. 2005 in prep.). The early-type host-galaxy is dominated by a post-starburst stellar population of only about 0.2 Gyr old. The offset in time between the early stage of the merger (roughly 2 Gyr ago as concluded from the \HI\ distribution) and the ignition of the starburst in B2 0648+27 can be explained when the progenitor galaxies already contained a bulge component, that stabilized the central region from forming gravitational instabilities until the second encounter of the progenitor galaxies (see models by e.g. \cite{mih96,kap05,spr05}). We will also model the spectra of the other \HI-rich radio galaxies in order to look for (post)-starburst stellar populations, which could further constrain the age and properties of the merger event in these systems.

\subsection{Radio sources}
\label{sec:radio} 

The radio continuum structure of two of the \HI-rich radio galaxies in our sample (B2 0258+35 and B2 0648+27) has been studied in detail at VLBI-scale by \cite{gir05a}. They estimate that the radiative age of these sources is about 1 Myr, although there are also regions of likely much older continuum-emission, which suggests that these sources could have been confined for a significant time. Also for B2 1217+28 \cite{gir05b} argue that low-velocity jets in this very nearby radio galaxy cannot bore through the local ISM and escape. Nevertheless, we expect that the age of these three radio sources is still at least a few orders of magnitude smaller than the timescale of the merger events. We expect that the same is true for the other radio sources in our sample, since the typical lifetime of FR-I radio sources is generally believed to be only several $\times 10^{7}$ yr (e.g. \cite{par02}). The other compact sources in the \HI-rich radio galaxies can even be much younger than this, although it is uncertain whether these sources are compact because they are young, or because they have been confined for a significant time.

\begin{figure}
\includegraphics[width=8.4cm]{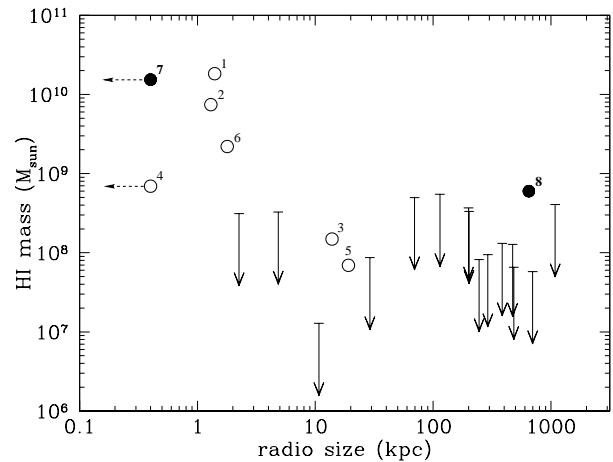}
\caption{Total \HI\ mass detected in emission plotted against the linear size of the radio sources. In case of non-detection an upper limit is given. The open circles and the arrows represent the sources that we observed, the closed circles are the two sources from the literature that we include for comparison. The numbers correspond to the sources as they are given in Table \ref{tab:hiradiogalaxies}.}
\label{fig:hisizeplot}
\end{figure}

A very interesting result from our study is shown in Figure \ref{fig:hisizeplot}, where we plot the mass of the extended \HI\ observed in our sample sources (or an upper limit in case of non-detection) against the size of the radio source. There is an apparent relation between the size of the radio source and the \HI\ content in nearby radio galaxies, in the sense that large amounts of \HI\ ($>$ 10$^{9}$ M$_{\odot}$) are only found around galaxies with a {\sl compact} radio source, while none of the galaxies with the more extended FR-I radio sources in our sample show such large amounts of \HI.

\section{Conclusions and discussion}
\label{sec:conclusions}

{\sl (1)} {\bf Detections of extended \HI\ associated with nearby radio galaxies show mainly large and fairly regular rotating disk-like structures}, which suggest that a major merger happened in these systems several Gyr ago. Since the typical lifetime of the radio source is likely much less than this, {\sl the current episode of radio-AGN activity must have started very late in the lifetime of the merger. Therefore the ignition of the radio-AGN is either significantly delayed, or it is not directly related to the merger event at all.}

\vspace{2mm}
\noindent {\sl (2)} {\bf In ${\bf 26 \%}$ of our sample sources we detect extended \HI.} This detection rate is in agreement with detection rates of \HI\ emission in samples of early-type galaxies that were not selected on the presence of AGN activity (see Morganti et al. in these proceedings), although the different detection limits of the various samples and the optical classification of the host-galaxies deserves further attention before making a firm statement about this. Nevertheless, together with the fact that the massive \HI\ disks that we find around our radio galaxies appear morphologically very similar to the \HI\ disks found around some non-active early-type galaxies (\cite{mor97,sad00,oos02}, 2003), this might be {\sl an indication that radio-AGN activity could occur at any point during the lifetime of any early-type galaxy.}

\vspace{2mm}
\noindent {\sl (3)} {\bf Large amounts of \HI\ ($> 10^{9}$ M$_{\odot}$) are only found around compact radio sources}, while none of the more extended FR-I radio sources in our sample show such large amounts of \HI. The nature of this apparent relation is uncertain. Possibly large amounts of gas, which could have been transported to the central region of the galaxy during the merger event, can confine the radio source (confinement has been proposed for B2 0258+35, B2 0648+27 and B2 1217+29 by Giroletti et al. 2005a,b). Alternatively, low-density \HI\ structures could be ionized when the radio jets break through the galaxy. Simulations by Bicknell $\&$ Sutherland (these proceedings) show that the propagating jets trigger a spherical bow-shock that expands through the host galaxy. If extended gas structures are present, this mechanism might offer an explanation on how they can be ionized or possibly even heated to X-ray temperatures. A third possibility is that the galaxies that contain large amounts of \HI\ and a compact radio source are fundamentally different from the galaxies with extended FR-I sources in our sample. More observations of the radio sources at VLBI-scales and of the host galaxies at wavelengths of ionized or X-ray gas are necessary to verify and explain {\sl the relation that we find between the size of the radio source and the \HI\ content of the host galaxy.}

\end{document}